**Efficient spin-to-charge interconversion in Weyl semimetal TaP at room temperature**

J. B. S. Mendes[1, *], R. O. Cunha[1], S. O. Ferreira[1], R. D. dos Reis[3], M. Schmidt[4], M. Nicklas[4], S. M. Rezende[2], and A. Azevedo[2]

[1]Departamento de Física, Universidade Federal de Viçosa, 36570-900, Viçosa, MG, Brazil
[2]Departamento de Física, Universidade Federal de Pernambuco, 50670-901, Recife, PE, Brazil
[3]Brazilian Synchrotron Light Laboratory (LNLS), Brazilian Center for Research in Energy and Materials (CNPEM), Campinas, SP, Brazil
[4]Max Planck Institute for Chemical Physics of Solids, Nöthnitzer Str. 40, 01187 Dresden, Germany

In this paper we present spin-to-charge current conversion properties in the Weyl semimetal TaP by means of the inverse Rashba-Edelstein effect (IREE) with the integration of this quantum material with the ferromagnetic metal Permalloy (Py=$Ni_{81}Fe_{19}$). The spin currents are generated in the Py layer by the spin pumping effect (SPE) from microwave-driven ferromagnetic resonance and are detected by a dc voltage along the TaP crystal, at room temperature. We observe a field-symmetric voltage signal without the contamination of asymmetrical lines due to spin rectification effects observed in studies using metallic ferromagnets. The observed voltage is attributed to spin-to-charge current conversion based on the IREE, made possible by the spin-orbit coupling induced intrinsically by the bulk band structure of Weyl semimetals. The measured IREE coefficient $\lambda_{IREE} = (0.30 \pm 0.01)\ nm$ is two orders of magnitude larger than in graphene and is comparable to or larger than the values reported for some metallic interfaces and for several topological insulators.

*Corresponding author: Joaquim B. S. Mendes, joaquim.mendes@ufv.br

The development of spintronic devices based on the electrical manipulation of magnetization, has been fueled by the discovery of various topological materials [1-3]. At first, it was thought that the presence of magnetic materials (or strong magnetic fields) could be a problem in topological materials, as it would lead to time-reversal symmetry breaking. This would make surface states no longer be topologically protected, leading to electron backscattering and localization, thus causing dissipation in transport. On the other hand, effects that are even more interesting have been discovered when the T-symmetry is broken by combining magnetic, topological and large spin orbit coupling (SOC) materials. These developments gave rise to the area of topological spintronics that encompasses applications in spin-orbit torque (SOT) for low-energy switching, skyrmion motion for racetrack memory, spin charge conversion for data processing, etc. [4-6]. Among the topological materials, the Weyl semimetals (WSM), that can be viewed as a three-dimensional analogue of graphene, have pairs of nondegenerate bands that intersect at gapless points, called Weyl nodes, around which they exhibit a linear dispersion relation (Weyl cone). The electrons near these points satisfy the Weyl equation, which is a two-component analog to the Dirac equation. They behave like massless particles and exhibit a variety of exotic properties such as the chiral anomaly as well as the presence of open Fermi surfaces and extraordinary transport properties. In contrast to topological insulators, WSM are conductors and have similar spin-momentum locking in both surface and bulk states. WSMs feature Dirac-like cones in their bulk near the Fermi energy (FE) through nodal lines or points (Weyl points) on or near some crystalline planes of symmetry, such as mirror planes. Some nodal lines and points are not protected by crystalline symmetry, so these are gapped out by the SOC that is intrinsic in these materials. The sign of the spin Berry curvature (SBC) is opposite on either side of the gap, which cannot be cancelled out if the Fermi level is in or very close to the gap. This gapped nodal lines/points generate large intrinsic spin Hall conductivity (SHC) which is proportional to the integration of the SBC of the occupied bands below. These exotic properties make them interesting for both basic science and technological applications. Therefore, WSM are considered to have a huge potential for ultralow-power spintronic devices with an efficient conversion of spin-to-charge current, i.e., a large spin Hall effect (SHE) and/or Rashba-Edelstein effect (REE) at room temperature [7–11].

Among the WSM, the TaP compound belong to the TaAs family, with the other three members TaAs, NbP and NbAs, which represents the first experimental realization of WSMs. This family of compounds crystallize in non-centrosymmetric lattice with mirror symmetry in the (100) and (010) directions. The broken symmetry of the crystal gives rise to 12 Weyl-point pairs, 8 pairs called W1 sitting at the top and bottom of the Brillouin zone and 4 pairs called W2 at half height at zone boundaries [11-21]. Each pair is separated by one mirror plane. Previous work by Arnold et al. [12], combined quantum oscillation data with ab-initio calculation to reveal the picture of the Fermi surface of the TaP compound. Their work demonstrates that in TaP the Weyl points sit energetically slightly away from the Fermi energy, with W1 points at -41 meV and W2 at 13 meV. This offset makes the W1(W2) Weyl points sit inside small electron(hole) pockets and weakens the Weyl fermion characteristics of the quasiparticles. More interesting, the fact that the Fermi surface has similarly sized pockets of electrons and holes makes the TaP material almost compensated with the electron density close to the hole density ($2\times10^{19}$ cm$^{-3}$). This balance between the charge carriers in combination with the exceptionally high carrier mobilities (~$10^5$ cm$^2$/Vs) leads TaP to present an extremely high and non-saturating magnetoresistance with values of the order of $10^6$ % at 2 K. In addition to the topological Weyl features in these semimetals, trivial spin-polarized Fermi arc surface states are also shown to exist at the Fermi level between the electron and hole pockets at room temperature. As the interconversion between charge and spin currents is key for integrating electronics with spintronics, WSM are strongly considered for the development of spin-based devices.

In this paper, we report the observation of efficient spin-to-charge current conversion by means of the IREE in Weyl semimetal TaP at room temperature. The spin currents are generated by microwave-driven ferromagnetic resonance (FMR) spin pumping in a ferromagnetic film of Permalloy (Py = Ni$_{81}$Fe$_{19}$). The experiments were carried out with heterostructures as illustrated in Fig. 1(a). Several samples of TaP/Py were prepared using high-quality single crystals of TaP grown via a chemical vapor transport reaction and with lateral dimensions of 2.0 x 1.0 mm$^2$. Scanning electron microscopy (SEM) image for a typical crystal obtained by this technique can be seen in Fig.1(b). The NiFe film was deposited on a flat area of the TaP crystal, as indicated by the blue dashed rectangle. More details on the TaP-crystal preparation and characterization can be found in Refs. [8,9]. To verify the high quality of our sample we performed longitudinal resistivity $\rho_{xx}$ measurements as a function of temperature, as depicted in Fig. 1(c). The inset in this figure shows an image of the four-point probes made exclusively for this measurement. We find a good agreement with previously reported data [22], with the sample displaying a metallic behavior with a resistivity ratio $\rho_{xx}$(300 K)/$\rho_{xx}$(2 K) ≈ 132.

The Py film (50 nm) was grown by DC sputtering and then two thin metal electrodes were fixed to the TaP-crystal by means of silver ink, as illustrated in Fig. 1(a). Figure 1(d) shows the *I-V* curve for samples used in this work, exhibiting the ohmic electrical contact between the Ag electrodes and the TaP. We observe that the resistance of the TaP-crystal is ~1 Ω, which is much lower than the electrical resistance of the Py film (~50 Ω). Therefore, the current flowing through the device mostly flows through the TaP, which minimizes any unwanted galvano-magnetic effects of the Py. Thus, the electrical detection of the spin-current to charge-current conversion comes from the TaP, by either bulk or surface states. Figure 1(e) shows Raman scattering measurement of TaP crystal at room temperature, obtained with a Renishaw In Via micro-Raman system equipped with a diode laser of wavelength 633 nm. The measurements were performed with all available laser on flat sample surface regions using a long distance 50x objective lens, resulting in a laser spot having diameter around 1 μm in the focal plane and the power at the sample was smaller than 8 mW. The spectrum of Fig. 1(e) shows the characteristic peaks related to the energy modes E(1) (~141 cm$^{-1}$), E(2) (~337 cm$^{-1}$), A$_1$(~373 cm$^{-1}$) and B$_1$(2) (410 cm$^{-1}$) for TaP [23]. The structure characterization was assessed with a Bruker D8 Discover diffractometer equipped with the Cu K-α radiation (λ=1.5418Å). Fig. 1(f) shows the out-of-plane X-ray diffraction (XRD) $\theta$-$2\theta$ scan pattern of the TaP sample, exhibiting reflections associated with the (004), (008), (105) and (200) crystal planes, and also indicating that no impurity phases precipitated.

In the spin pumping experiments, the TaP/FM bilayer sample is under action of a weak radio-frequency magnetic field ($h_{rf}$) applied perpendicular to a dc magnetic field in the FMR configuration. The precessing magnetization $\vec{M}$ in the FM layer generates a spin current density $\vec{J}_S$, that flows perpendicularly to the TaP/FM interface with transverse spin polarization $\hat{\sigma}$, which is given by $\vec{J}_S = \left(\hbar g_{eff}^{\uparrow\downarrow}/4\pi M_s^2\right)\left(\vec{M}(t) \times \partial\vec{M}(t)/\partial t\right)$. Here, $M_s$ is the saturation magnetization, and $g_{eff}^{\uparrow\downarrow}$ is the real part of the effective spin mixing conductance at the interface that takes into account the spin-pumped and back-flow spin currents [24]. Writing the magnetization as $\vec{M}(t) = M_z\hat{z} + (m_x\hat{x} + m_y\hat{y})e^{i\omega t}$, where $m_x, m_y \ll M_z$, the DC spin current density pumped at the Py/TaP interface is given by:

$$J_S(0) = \frac{\hbar \omega g_{eff}^{\uparrow\downarrow} p}{16\pi}\left(\frac{h_{rf}}{\Delta H}\right)^2 L(H - H_r), \qquad (1)$$

where $L(H - H_r) = \Delta H^2/[(H - H_r)^2 + \Delta H^2]$ is the Lorentzian function, $p$ is the ellipticity factor for the precession cone $p = 4\omega\,(H_r + 4\pi M_{eff})/\gamma\,(2H_r + 4\pi M_{eff})^2$, and $\Delta H$ is the FMR linewidth (half-width half-maximum). It is important to mention that the equation (1) is in units of (angular moment per (area x time)). The pure spin current that flows through the Py/TaP interface is formed by charge carriers with opposite spins moving in opposite directions, diffuses into the TaP creating a spin accumulation given by $J_S(y) = J_S(0)\{\sinh[(t_{TaP} - y)/\lambda_S]/\sinh(t_{TaP}/\lambda_S)\}$, where $t_{TaP}$ and $\lambda_S$ are the thickness and spin diffusion length of the TaP layer. Due to the inverse SHE (ISHE), part of this spin current is transformed in a transverse charge current density given by $\vec{J}_C = \theta_{SH}(2e/\hbar)\vec{J}_C \times \hat{\sigma}$, where $\theta_{SH}$ is the spin Hall angle (which measures the spin to charge current conversion efficiency) and $\hat{\sigma}$ is the spin polarization. By integrating the charge current density along x and y (see inset in Fig 1(a)), we obtain the spin pumping current, $I_{SP} = V_{SP}/R = w\lambda_S \theta_{SH}(2e/\hbar)\tanh(t_{TaP}/2\lambda_S) J_S(0)$, where $w$ is the width of the TaP layer. The above analysis is valid for the spin-charge conversion that occurs due to ISHE, which is effective in bulk states, but can also occur in surface states by the IREE. By investigating these two independent phenomena (ISHE and IREE), it is possible to extract material parameters from the measurements of the FMR absorption and the spin-pumping voltage.

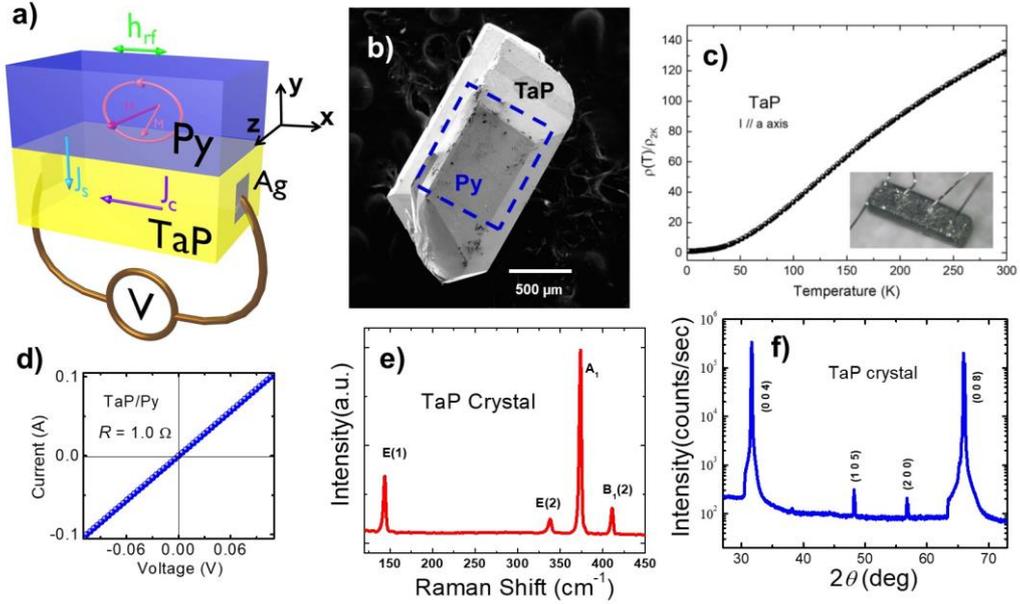

**Figure 1**. (Color online) (a) Illustration of the TaP/Py structures and the electrodes used to measure the DC voltage due to the spin pumping effect. (b) SEM image for a typical TaP crystal used in this work. The Py film was deposited on the rectangle blue dashed region of the TaP crystal. (c) Temperature dependent of the normalized electrical resistivity $\rho(T)/\rho$ (2K) in zero magnetic field. The inset shows a picture of the single crystal used for the electrical transport characterization. (d) *I-V* curve showing the ohmic behavior of the contacts between the electrodes and the WSM. (e) Raman spectrum of TaP crystal at room temperature which shows the characteristic peaks related to energy modes E(1), E(2), A1 and B1(2). (f) XRD spectrum of the bulk TaP measured at room temperature.

For the FMR and spin pumping experiments, the sample of Fig. 1(a) was mounted on the tip of a polyvinyl chloride (PVC) rod and inserted through a hole drilled in the center of the back wall of a rectangular microwave cavity operating in the transverse electric ($TE_{102}$) mode, at a frequency of 9.4 GHz with a $Q$ factor of 2000. The sample is slightly inserted into the cavity in the plane of the back wall, in a position of maximum rf magnetic field and zero electric field to avoid the generation of galvanomagnetic effects driven by the electric field. The microwave cavity is placed between the poles of an electromagnet so that the sample can be rotated while maintaining the static and rf fields in the sample plane and perpendicular to each other. With this configuration we can investigate the angular dependence of the FMR absorption spectra. Field scan spectra of the derivative $dP/dH$ of the microwave absorption are obtained by modulating the dc field with a weak ac field at 1.2 kHz that is used as the reference for lock-in detection. Figure 2(b) shows the FMR absorption spectrum of the Py layer in contact with the TaP film measured with incident microwave power of 24 mW. The FMR line has the shape of a Lorentzian derivative with peak-to-peak linewidth of 50.8 Oe, corresponding to a half-width at half-maximum (HWHM) linewidth of $\Delta H_{TaP/Py} = 44$ Oe. As shown in the Fig. 2(a), an identical Py layer deposited on a Si substrate has linewidth $\Delta H_{Py} = 23$ Oe, showing that the contact of the TaP layer produces an additional damping due to the spin pumping process [25,26] of $\delta H = 21$ Oe.

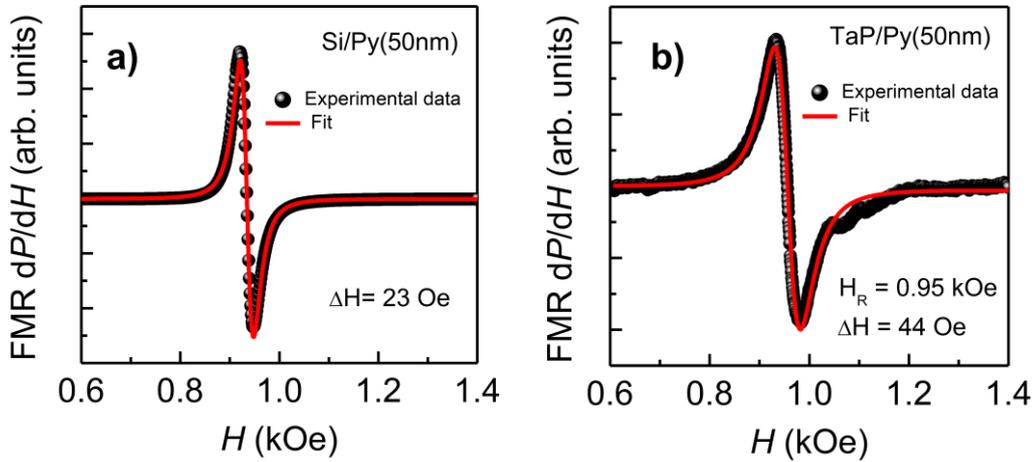

**Figure 2**. (Color online) Panels (a) and (b) show the field scan FMR microwave absorption derivative spectra at a frequency 9.4 GHz of two samples of 50 nm thick NiFe film in atomic contact with Si and TaP as indicated, with the magnetic field applied in the film plane, normal to the long dimension.

Similar effect has been observed in bilayers formed by Py in atomic contact with different materials: normal metals/Py [27-31], semiconductors/Py [32-37], graphene/Py [38-40], topological insulators/Py [41-43], etc. Since the magnetic anisotropy of Py is small, the field of resonance $H_R$ is related to the microwave frequency by the Kittel equation $f = \gamma (H_R)^{1/2} (H_R + 4\pi M_{eff})^{1/2}$, where $\gamma$ is the gyromagnetic ratio and $4\pi M_{eff}$ is the effective magnetization, which is smaller than the saturation magnetization due to the effect of the perpendicular anisotropy in thin films. Using the measured $H_R = 0.95$ kOe and $\gamma = 2.94$ GHz/kOe, corresponding to a *g*-factor for Py of 2.1, we obtain $4\pi M_{eff} = 9.81$ kG, which is consistent with values for a 50 nm thick Py layer at room temperature [28, 44].

Figure 3(a) shows the field scan *dc* voltage measured directly with a nanovoltmeter connected by copper wires to the electrodes, for a microwave power of 150 mW, for angles of the in-plane field, $\phi = 0°$ and $\phi = 180°$. The voltage is positive for $\phi = 0°$, changes sign with inversion of the field, and vanishes for the field along the sample strip $\phi = 90°$. The asymmetry between the positive and negative peaks amplitudes is similar to that observed in other bilayer systems and can be attributed to a thermoelectric effect [45]. In turn, the magnetic field dependence of $V_{SP}(H)$, measured between the electrodes exhibits an asymmetric Lorentzian line shape around the FMR field $H_R$. Thus, $V_{SP}(H)$ can be described by the sum of two components, $V_{SP}(H) = V_{sym} L(H - H_R) + V_{asym} D(H - H_R)$, where $L(H - H_R)$ is the symmetric (absorptive) and $D(H - H_R)$ is the antisymmetric (dispersive) contributions, respectively. Lorentzian derivative centered around the FMR field $H_R$, where the terms $V_{sym}$ and $V_{asym}$ denote the amplitudes of the symmetric and antisymmetric components. Note in Fig. 3(a) that the *H* dependence of $V_{SP}$ exhibits a symmetric Lorentzian line shape with a negligible antisymmetric component. This suggests that the contributions from galvanomagnetic or spin charge pumping effects [28,46], generated by the Py layer itself can be neglected. On the other hand, the most important source for the symmetric component of the voltage is the conversion of the spin current produced by spin pumping into charge current in the WSM layer. The dependence of the spin pumping voltage as a function of the applied field, for different rf powers ($P_{rf}$), is shown in Fig. 3(b), for $\phi = 0°$. Figure 3(c) shows that the dependence of the maximum value of the spin pumping voltage ($V_{SP}^{peak}$) on the rf power is in the linear regime of excitation, showing that nonlinear effects are not being excited in the power range of the experiments.

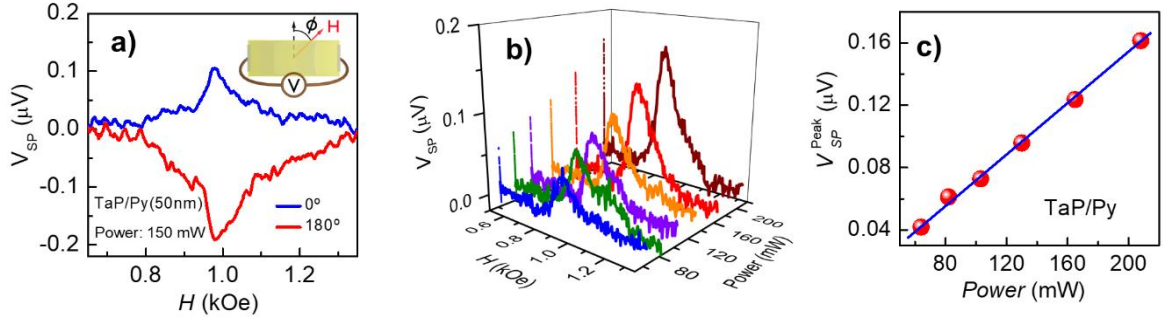

**Figure 3**. (Color online) (a) Field scan of the spin pumping voltage measured in TaP/Py at different in-plane field angles as illustrated in the inset, with an incident microwave power of 150mW. (b) Field scans of $|V_{SP}|$ for several values of the incident microwave power, and (c) peak voltage value as function of the rf excitation power measured on the TaP/Py sample.

To compare the spin pumping response of the TaP sample with the responses obtained for other topological materials (TM), we use the spin pumping current $I_{SP} = V_{SP}/R$, where $R$ is the electric resistance between the two electrodes. $I_{SP}$ is a more significant physical parameter because $V_{SP}$ is proportional to the resistance, which a parameter dependent on electric contact artifacts. Figure 4 (a) shows the field dependence of $I_{SP}$ for four different bilayers TM/Py, where the used TMs are: single layer of graphene (SLG), $Bi_2Se_3$(4 nm) and $(Bi,Sb)_2Te_3$(4 nm), and Py is used as the spin current injector. The $I_{SP}$ field scans spectra for SLG/Py(12 nm), $Bi_2Se_3$(4 nm)/Py(12 nm) and $(Bi,Sb)_2Te_3$(4 nm)/Py(12 nm) obtained with the same experimental configuration are shown in Figs. 4(b), 4(c) and 4(d), respectively. While $I_{SP}$ peak is very weak in graphene and the values are equivalent for $Bi_2Se_3$ and $(Bi,Sb)_2Te_3$, it is much more stronger for TaP, confirming that TaP has an efficient spin-to-charge interconversion, as seen in Fig. 4(a). The solid lines were obtained by adding the symmetric and antisymmetric contributions. According to equation (1) the symmetric component of $I_{SP}$ is due to the spin-charge conversion process. We can define a figure of merit to quantify the process as $\xi(A/B) = I^A_{sym}/I^B_{sym}$, where the letters A and B represent material A and material B, respectively. Thus, $\xi(TaP/SLG) \cong 40.5$, $\xi(TaP/Bi_2Se_3) \cong 7.6$ and $\xi(TaP/(Bi,Sb)_2Se_3) \cong 7.6$. Surprisingly, TaP is much more efficient to convert spin-current into charge-current than other topological materials. This large efficiency can be attribute to the unique properties of the Weyl semimetal TaP. Before to speculate about the prevalent mechanism acting in the spin-to-charge conversion process in TaP, we can calculate the effective spin mixing conductance, which is given by $g^{\uparrow\downarrow}_{eff} = (4\pi M_S t_{Py}/\hbar\omega)\delta H$ [31,40,43]. As $4\pi M_S \cong 11$ kG, $t_{Py} = 50$ nm, $\omega/2\pi = 9.4$ GHz, and the additional linewidth broadening measured in Py due to the contact with the TaP layer is $\delta H = (\Delta H_{TaP/Py} - \Delta H_{Py}) = 21$ Oe, we obtain $g^{\uparrow\downarrow}_{eff} = 1.4 \times 10^{19}$ m$^{-2}$. This value is similar to the one for Py/Pt interfaces, demonstrating an efficient spin transfer in TaP/Py heterostructures [28,30,31,37].

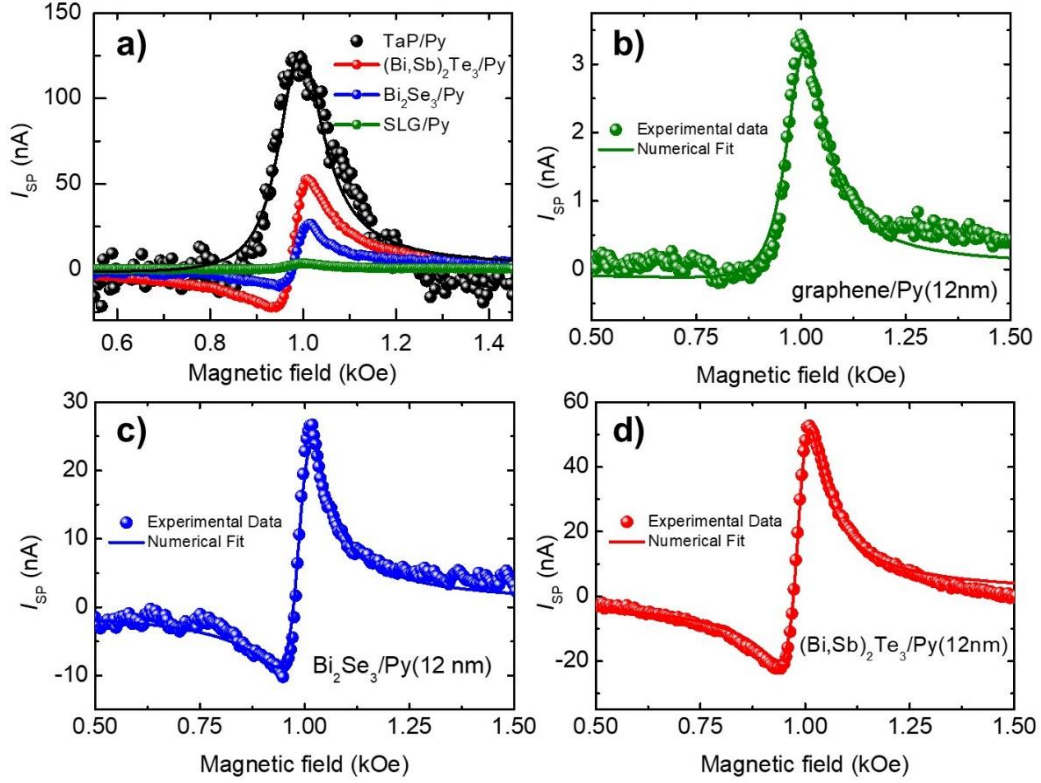

**Figure 4**. (Color online) (a) Field dependence of $I_{SP}$ for TaP/Py sample at $\phi = 0°$ compared to other topological hybrid nanostructures. The dc electrical current was obtained with an incident rf power of 165 mW and frequency of 9.4 GHz for all samples. The field scan spectra of $I_{SP}$ in $\phi = 0°$ and for the bilayers (b) SLG/Py, (c) $Bi_2Se_3$/Py and (d) $(Bi,Sb)_2Te_3$/Py obtained with the same power 165 mW, and the decomposition of the symmetric components of $I_{SP}$ obtained by fitting the experimental data.

To understand the role played by the boundary and volume states in the spin-to-charge current conversion process in TaP we can calculate the spin diffusion length assuming that the prevalent mechanism is due to the ISHE or IREE. First, let us assume that the predominant mechanism is due to ISHE. As the TaP sample is sufficiently thick, i.e., $t_{TaP} \gg \lambda_S \rightarrow \tanh(t_{TaP}/2\lambda_S) \cong 1$ and considering the peak of the SP current ($I_{SP}$) at $H = H_R$, i.e., $L(H - H_R) = 1$, the peak value is given by $I_{SP} = (1/4)ewfg_{eff}^{\uparrow\downarrow}p(\lambda_S\theta_{SH})(h_{rf}/\Delta H)^2$, where $w = 10^{-3}$ m, $e = 1.6 \times 10^{-19}$ C, $f = 9.4$ GHz, $g_{eff}^{\uparrow\downarrow} = 1.4 \times 10^{19}$ m$^{-2}$, $h_{rf} = 1.776\sqrt{0.15} = 0.687$ Oe, $p = 0.31$, $\Delta H = \Delta H_{TaP/Py} = 44$ Oe and $I_{SP} = 130$ nA we obtain, $\Theta_{SP} = \lambda_S\theta_{SH} \cong$ 0.31 nm. It is also useful to define the spin Hall angle according to $\theta_{SH} = (2e/\hbar)(\sigma_{SH}/\sigma)$, where $\sigma_{SH}$ and $\sigma$ are the spin Hall conductivity and the charge conductivity of the TaP, respectively. Using the average values for the conductivities $\sigma$ and $\sigma_{SH}$ that were previously reported [12,47], we obtain $\theta_{SH} = 0.04 \pm 0.01$ (or $\theta_{SH} \cong 4\%$), which is about three times greater than the one measured in WTe$_2$/ferromagnet bilayers [48]. In turn, we also obtained the spin diffusion length $\lambda_S(TaP) = (8 \pm 1)$ nm, which is similar to the values reported for the WTe$_2$ [49]. Therefore, our results indicate that the volume states are playing a crucial role in the spin-to-charge current conversion.

Now, let us assume that the prevalent mechanism is IREE. By using equation (1), the amplitude of the microwave field in Eq. (1), is related to the incident power $P_i$, in watt, by $h_{rf} = 1.776 (P_i)^{1/2}$ [43], calculated for a X-band rectangular microwave cavity operating in the $TE_{102}$ mode with $Q$ factor of 2000, at a frequency of 9.4 GHz. Using these values, we obtain for $P_i = 150$ mW, $H = H_R$, the spin current density at the interface produced by the FMR spin pumping $J_S = 2.2 \times 10^6$ A/m$^2$. The charge current density due to the conversion from the spin current by the IREE, given by $j_C = V_{IREE}^{peak}/(wR_S)$, corresponding to the measured voltage of $V_{IREE}^{peak} = 0.2\,\mu$V, considering that the shunt resistance $R_S = 1.0$ Ω, and $w = 1.0$ mm, is $j_C = 2.0 \times 10^{-4}$ A/m. We consider that the 3D spin current density ($J_S$) generated by the spin pumping process in Py, flows into the TaP layer and is converted by the IREE into a lateral charge current with 2D density $j_C$. This process is quantified by the IREE coefficient $\lambda_{IREE}$ defined through the relation $j_C = (2e/\hbar)\lambda_{IREE} J_S$. Note that the IREE coefficient has dimension of length and is proportional to the Rashba coefficient, and hence to the magnitude of the SOC [43]. Using the relation $V_{IREE} = R_S w j_C$ in Eq. (1) we obtain an expression for the IREE coefficient in terms of the measured voltage peak value

$$\lambda_{IREE} = \frac{4V_{IREE}}{R_S e w f g_{eff}^{\uparrow\downarrow} p_{xz} (h_{rf}/\Delta H)^2}. \qquad (2)$$

Using the physical quantities for the bilayer TaP/Py (50 nm) obtained from the spin pumping measurements, the IREE coefficient value is $\lambda_{IREE} = (0.30 \pm 0.01)$ nm. The error bar has been incorporated in $\lambda_{IREE}$ by taking into account the variation of $V_{SP}$ measured at $\phi = 0°$ and 180°. This value is two orders of magnitude larger than in graphene [40,50] and comparable or larger than the values reported for several topological insulators [43, 51-56].

In summary, we have demonstrated the conversion of a spin current into a charge current in the Weyl semimetal TaP at room temperature, which is attributed to the inverse Rashba-Edelstein effect made possible by spin momentum locking in the electron Fermi contours due to the Rashba field. The spin currents were generated in a layer of permalloy by spin pumping effect (SPE). In this case, we have used microwave-driven ferromagnetic resonance of the Py film to generate a spin current that is injected into the TaP crystal in direct contact with Py. The results of the measurements yield a value for the IREE coefficient, $\lambda_{IREE} = (0.30 \pm 0.01)$ nm, which is two orders of magnitude larger than in graphene and is comparable to or larger than the values reported for some metallic interfaces and for several topological insulators.

**Supporting Information**

Supporting Information is available from the Wiley Online Library or from the authors.

**Acknowledgements**

This research was supported by Conselho Nacional de Desenvolvimento Científico e Tecnológico (CNPq), Coordenação de Aperfeiçoamento de Pessoal de Nível Superior (CAPES), Financiadora de Estudos e Projetos (FINEP), Fundação de Amparo à Ciência e Tecnologia do Estado de Pernambuco (FACEPE), Fundação de Amparo à Pesquisa do Estado de São Paulo (FAPESP- Grant: 2018/00823-0), Fundação de Amparo à Pesquisa do Estado de Minas Gerais (FAPEMIG) - Rede de Pesquisa em Materiais 2D and Rede de Nanomagnetismo. R.D. dos Reis acknowledges the Max Planck Society under the auspices of the Max Planck Partner Group R. D. dos Reis of the MPI for Chemical Physics of Solids, Dresden, Germany.

**Conflict of Interest**

The authors declare no conflict of interest.

**Data Availability Statement**

The data that support the findings of this study are available from the corresponding author upon reasonable request.

**Keywords**

Spin-charge conversion, ferromagnetic resonance, spin pumping, topological Weyl semimetal